\documentstyle[12pt]{article}
\newcommand{\be}{\begin{equation}}
\newcommand{\bea}{\begin{eqnarray}}
\newcommand{\eea}{\end{eqnarray}}
\newcommand{\ba}{\begin{array}}
\newcommand{\ea}{\end{array}}
\newcommand{\ee}{\end{equation}}

\expandafter\ifx\csname mathbbm\endcsname\relax

\else

\fi
\begin{document}
\begin{titlepage}
\hfill
\vbox{
    \halign{#\hfil         \cr
           ITFA-2001-17 \cr
           hep-th/0105153  \cr
           } 
      }  
\vspace*{20mm}
\begin{center}
{\Large {\bf (De)Constructing Dimensions and Non-commutative Geometry}\\ } 

\vspace*{15mm}
\vspace*{1mm}
{Mohsen Alishahiha }\\

\vspace*{1cm} 

{\it Institute for Theoretical Physics, University of Amsterdam,\\
Valckenierstraat 65, 1018 XE Amsterdam, The Netherlands}\\

\vspace*{1cm}
\end{center}

\begin{abstract}
In this paper the model considered by Arkani-Hamed, Cohen and Georgi in the
context of (de)constructing dimensions has been studied
by making use of non-commutative geometry. The non-commutative geometry
provides a natural framework to study this model with or without gravity. 
\end{abstract}
\vskip 6cm

\end{titlepage}

\newpage

{\em We are all agreed that your theory is crazy. The question which
divides us is whether it is crazy enough.}

$\;\;\;\;\;\;\;\;\;\;\;\;\;\;\;\;\;\;\;\;\;\;\;\;\;\;\;\;\;\;\;\;
\;\;\;\;\;\;\;\;\;\;\;\;\;\;\;\;\;\;\;\;\;\;\;\;\;\;\;\;\;\;\;\;
\;\;\;\;\;\;\;\;\;\;\;\;\;\;\;\;\;\;\;\;\;
$ - Niels Bohr

\section{Introduction}
Although it seems we live in a four-dimensional world, it has been suspected 
that at short distance, shorter than it has been probed yet, the best 
description of our world could be provided by a theory with more than 
four-dimensional spacetime. The simplest example could be that with four 
extended 
dimensions plus one compact dimension. In this case, at the distance much 
bigger than the size of the radius of the compact direction, the theory looks 
like a four-dimensional theory, 
while for the distance comparable to the size of the compact direction, the
effects of the five-dimensional theory will be appeared.
A generic feature of the theories with a small compact dimension is that, the
higher dimensional theory appears at high energy (UV limit), while the lower 
dimension description emerges at large distance (IR limit).

In an reversal picture, the authors in\cite{ACG} considered a theory in 
which the higher dimensional description is given in the IR limit. In fact,
in UV limit, where theories with higher than four dimensions are going to be
problematic, the theory is well described in terms of a four dimensional theory
, and actually in the extreme UV, the theory is perfectly four dimensional !
By making use of this strategy the authors in \cite{ACG} suggested a way
for a UV completions of the higher dimensional field theories. 

In fact, this is a generic property of the field theory on a four
 extended dimensions plus some {\it discrete} extra dimensions. Here
we shall only consider
 one extra discrete dimension. Suppose $a$ be a length 
scale of
the discrete dimension and $l$ the scale of the four-dimensional 
theory. In general, the different phases of the theory are parameterized by 
a dimensionless parameter $\zeta=a/l$. In the limit $\zeta \gg 1$
(UV) the theory is four dimensional and for $\zeta\ll 1$ (IR) the description
of the theory is given by a five dimensional field theory. 

If we think about this theory as a five-dimensional field theory 
{\it latticized} in one dimension, the parameter $a$ plays the role of the
lattice size and IR limit is the limit in which $a\rightarrow 0$ where
 we recover
five continuous dimensional theory.

The model which has been studied in \cite{ACG} is a quiver ( ``moose'' in their
notation) model with $SU(n)^N$ gauge theory coupled to $N$ non-linear sigma
model which can be obtained by starting with a quiver model with
gauge group $\prod_{i=1}^{N}SU_i(n)\times SU_i(m)\times 
SU_{i+1}(n)$ where $i=1$ is periodically identified with $i=N+1$. There are also
fermions transforming bi-linearly under nearest-neighbor pairs of the
gauge transformation. Suppose $\Lambda_n$ and $\Lambda_m$ be the energy 
scales of the gauge groups $SU(n)$ and $SU(m)$, respectively. For the limit
where $\Lambda_m \gg \Lambda_n$ and in the energy of order $\Lambda_m$, 
the $SU(m)$
groups become strong, causing the fermions to condense in pairs. The
confining strong interactions also produce a spectrum of hadrons
with masses on the order of $\Lambda_m$. Therefore, below the scale 
$\Lambda_m$ the theory can be described as a gauge theory with gauge
group $SU(n)^N$ coupled to a $N$ non-linear sigma model.\footnote{
An infinite arrays of gauge theories has also been studied in \cite{HS}, where
an infinite number of gauge theories are linked by scalars
to get an infinite tower of massive vector mesons  (``hadrons'') with a small
coupling only for the single zero mass photon. It seems that this theory is
in the same class as one considered in \cite{ACG}.}

One could think about this theory as a latticization of a five-dimensional
gauge theory with lattice size $a\sim (g \Lambda_m)^{-1}$ where $g$ is the
gauge coupling of the gauge group $SU(n)$. While the theory is a 
four-dimensional gauge theory in UV, at large distance it turns out to be
a five-dimensional gauge theory compactified on a circle of circumference
$R=Na$. In this sense, the extra dimension has been generated dynamically
\cite{ACG}. 
 
In general, having a manifold with a discrete dimension would technically 
cause a problem,
as the classical notion of the differential geometry fails for such a manifold.
In particular, the notion of curvature and torsion, which we need if we wish
to add gravity in the game, are not well-defined for such a manifold in terms of 
the classical differential
geometry. Fortunately, there is a generalization of the classical geometry 
for such a manifold in the context of, so called,
non-commutative geometry \cite{C}. 
Non-commutative geometry provides  a strong tools to 
study a manifold with discrete dimension, and in fact, using non-commutative
geometry one can study manifolds which could have no well-known
geometrical picture. Indeed, a well-known example in the non-commutative 
geometry is what we are interested in, i.e. a four-dimensional manifold
times a discrete set of $N$ points which altogether can be thought as a
five-dimensional spacetime with four extended dimensions and one discrete
dimension.

This is the aim of this note to reconsider the model studied in \cite{ACG}
in the framework of the non-commutative geometry.\footnote{
A possible connection between the non-commutative geometry and
 those theories with extra dimensions, like 
the Randall-Sundrum model \cite{RS}, has also been observed in\cite{LMM}.}
 An advantage using this point of 
view is that, by making use of the non-commutative geometry one can easily add
the gravity in the theory, much similar to what we would like to do for 
Yang-Mills sector. Indeed, one could have both Yang-Mills and
gravity sectors in the same time. In particular, the dynamically generating 
dimension procedure \cite{ACG} works for both gravity and  Yang-Mills sectors 
in the same way.

This paper is organized as following: In the section 2, we shall present a
brief review of the non-commutative geometry. In section 3, we will consider 
a non-commutative space which is given by a four-dimensional flat manifold times
a discrete set of $N$ points which can be thought as $N$ parallel 
four dimensional layers (sheets). Then we construct a gauge theory in this
space which corresponds to a four-dimensional gauge theory with
gauge group $SU(n)^N$ coupled to $N$ charged scalers which in general have
a quartic potential. This model is very similar to that considered in 
\cite{ACG}. In section 4, we will study the gravity in this space.
Finally, we shall give our conclusions in the section 5.

We note, however, that the formulas given in this paper have already been
presented in the literature, mostly, for the case of $N=2$ in the context of 
the 
non-commutative geometry applied to standard model of the particle physics.
The thing is new in this paper would be the generalization of those
results for $N$-points space, especially for the gravity sector, plus a
new interpretation in the context of the dynamically generating dimension.  

In this paper we use the following convention: the signature of the metric is
$(-, +, +,\cdots, +)$ and $\{\gamma^{\mu},\gamma^{\nu}\}=2g^{\mu \nu},
\; [\gamma^{\mu},\gamma^{\nu}]=2\gamma^{\mu\nu},\;
(\gamma^5)^2=1$.

\section{ A Brief Review of Non-commutative Geometry} 

There is a well-known theorem due to Gelfand and Naimark that 
a smooth manifold, $\cal M$, can be studied by analyzing the commutative
algebra ${\cal C}^\infty({\cal M})$ of smooth functions
defined on ${\cal M}$. In other words, the smooth manifold $\cal M$
can be reconstructed from the structure of ${\cal C}^\infty({\cal M})$.
The basic idea in the non-commutative geometry is 
how to define a compact, non-commutative space
in terms of a unital, non-commutative $*$-algebra ${\cal A}$ \cite{C}.

Given a unital, non-commutative $*$-algebra ${\cal A}$ one can define
the universal, differential algebra $\Omega({\cal A})$ for the 
non-commutative space.
For this purpose, assume $d$ to be an
abstract differential operator which acts on elements of ${\cal A}$ and
satisfies  the Leibniz rule, with $ d1=0, d^2 a=0$ where $1,a \in {\cal A}$.
Therefore we have
\be
\Omega({\cal A})=\bigoplus_{n=0}^{\infty} \Omega^n({\cal A})
\ee
with $\Omega^0({\cal A})={\cal A}$ and
\be
\Omega^n({\cal A})=\left\{\sum_{i} \,a_0^i\, da_1^i\, ... \,  da_n^i \;|\;
a_j^i \in {\cal A}, \forall i,j \right\},\;\;\;\;\;n=1,2,\cdots.
\ee
In fact $\Omega^n({\cal A})$ plays the role of space of $n$-form
in the non-commutative geometry.

The next ingredient which plays an important role in the differential structure
of the non-commutative geometry is the notion of Dirac $K$-cycle for $\cal A$.
The Dirac $K$-cycle is defined by a doublet $({\cal H},D)$ where $\cal H$ is
a Hilbert space and $D$ a selfadjoint operator on $\cal H$ (Dirac operator),
together with an involutive representation, $\pi$, of $\cal A$ on $\cal H$
\be
\pi:{\cal A}\rightarrow B({\cal H}),\;\;\;\;\;\;\pi(a^*)=\pi(a)^*,
\;\;\;\;\;\;\forall a\in {\cal A}
\ee
where $B({\cal H})$ is the algebra of the bounded operator on $\cal H$.

Given a Dirac $K$-cycle for $\cal A$, one can define an involutive 
representation of $\Omega({\cal A})$ on $\cal H$. This is provided by 
the map $\pi :\Omega({\cal A})\rightarrow B({\cal A})$ in such a way that,
for any element $\sum_{i} \,a_0^i\, da_1^i\, ... \,  da_n^i \in 
\Omega^n({\cal A}), n=1,2,\cdots$ we have
\be
\pi\left(\sum_{i} \,a_0^i\, da_1^i\, ... \,  da_n^i\right)=
\sum_{i} \pi(a^{0}_i)[D,\pi(a^1_i)]...[D,\pi(a^{p}_i)].
\ee
We note, however, that the representation $\pi$ is ambiguous \cite{CL}.
This can be seen as following. Suppose $\rho\in \Omega({\cal A})$ be a one
form. If $\pi (\rho)$ is set to zero, $\pi (d\rho)$ is not necessarily zero.
This fact leads us to define a set of auxiliary fields which appear because
of this ambiguity. By making use of the space of the auxiliary fields we can
correct the definition of the space of forms such that the ambiguity will be
removed. 

The space of the auxiliary fields is defined by ${\rm Aux}={\rm Ker}\pi+
d\; {\rm ker}\pi$, where
\bea
{\rm ker}\pi&=&\oplus_{n=0}^{\infty}\left\{\sum_{i} \,a_0^i\, da_1^i\, \cdots
 \,
da_n^i\;| \pi\left(\sum_{i} \,a_0^i\, da_1^i\, \cdots
 \,  da_n^i\right)=0\right\}\, ,
\cr
d\; {\rm ker}\pi&=&\oplus_{n=0}^{\infty}\left\{\sum_{i} \,da_0^i\, da_1^i\, 
\cdots \,
da_n^i\;| \pi\left(\sum_{i} \,a_0^i\, da_1^i\, \cdots
 \,  da_n^i\right)=0\right\}\, .
\eea
The space of the auxiliary fields is a two-sided ideal in $\Omega({\cal A})$
 and
this can be used to define the correct space of the forms as 
$\Omega_{D}({\cal A})=\Omega({\cal A})/{\rm Aux}$. Therefore for a element
$\sum_{i} \,a_0^i\, da_1^i\, ... \,da_n^i \in {\cal A}$,
\be
\left\{\sum_{i} \pi(a^{0}_i)[D,\pi(a^1_i)]...[D,\pi(a^{p}_i)] + \pi(\alpha)\;|
\; \alpha\in {\rm Aux}\right\}
\ee
represents an $n$-form (mod Aux) in $\Omega^n_{D}$ as an equivalence class of 
bounded operators on the Hilbert space $\cal H$.

The integral of a form $\beta\in \Omega({\cal A})$ 
over a non-commutative space $\cal A$ is defined by
\be
\int \beta={\rm Tr}_{\omega}\left(\pi(\beta) D^{-d}\right)\, ,
\ee
where ${\rm Tr}_{\omega}$ is the Dixmier trace and $d$ is the dimension of the
space represented by $\cal A$. The Dixmier trace is defined by
\be
{\rm Tr}_{\omega}(|T|)={\rm lim}_{\omega}\;\frac{1}{\log N}
\sum_{i}\mu_{i}(T)\, ,
\ee
where $T$ is a compact operator, and $\mu_i$ are the eigenvalues of $|T|$. 

One can also define a vector bundle over a non-commutative space $\cal A$, 
which is a free, projective $\cal A$-module. 
In fact a vector bundle, $E$, is defined by the vector space $\cal E$ of
its section which is going to be a free, projective, left $\cal A$-module. 
Here we are interested in the case ${\cal E}=\cal A$.

By making use of the structure of the non-commutative geometry, we will be
able to formalize a gauge theory on a non-commutative space. The procedure 
to define the Yang-Mills action is as following. As in the commutative case, we
would like to have a gauge connection and curvature which are one- and
two-form, respectively. Suppose $A\in \Omega^1({\cal A})$ be a gauge 
connection.
It can be expressed as 
\be
A=\sum_{\alpha} g_{\alpha}\;df_{\alpha}\, , 
\ee
with the condition $\sum_{\alpha} g_{\alpha}f_{\alpha}=1$. We need to impose
this condition in order to get correct gauge transformation under the unitary
gauge group $U({\cal A})=\{g\in {\cal A}\; |\; g^* g=1\}$ (see for example
\cite{CHAM2}). Of course this is no loss in generality, as the field
$\sum_{\alpha} g_{\alpha}f_{\alpha}$ is independent. As the usual case, the
curvature is defined by $F=dA+A^2$. Finally the Yang-Mills action is given by
\be
S_{YM}={1\over 8 }{\rm Tr}_{\omega}\left(\; \pi^2(F) D^{-4}\;\right)\, ,
\label{ACT}
\ee
here we assumed that the manifold represented by $\cal M$ is a four-dimensional
manifold. 
In the case we are interested in, the action (\ref{ACT}) reads (for example
see \cite{CHAM2})
\be
S_{YM}={1\over 8}\int d^4x\,\sqrt{{\rm det}(g)}{\rm Tr}\left(\pi^2(F)\right)\, ,
\ee
where $g$ is the metric and the trace, ${\rm Tr}$, is taken over both the
Clifford algebra and the matrix structure.

\section{N layers Model}

In this section we shall consider a non-commutative space which is
taken to be a product of a continuous four-dimensional manifold
times a discrete set of $N$ points.  Here, we assume that the
four-dimensional space is a flat space, and therefore this system could be 
thought as N parallel four dimensional layers. 
The proper algebra for this model is (we will only consider the case with
$N\geq 3$)
\begin{equation}
{\cal A}={\cal C}^\infty({\cal M}_4) \otimes (\,\, \oplus_{i=1}^{N}
M_n({I \kern -.6em C})\,\,).
\end{equation}
The Dirac operator can be chosen as follow
\be
D=\sum_{i=1}^{N}\left[\gamma^{\mu}\partial_{\mu}\;e_{i,i}+\gamma^{5}{K\over 
\sqrt{2}}\;
(e_{i,i+1}+e_{i,i-1})\right]\, ,
\label{DIC}
\ee
where $e_{i,j}$ is an $N\times N$ matrix with $(e_{i,j})_{ab}=\delta_{ia}
\delta_{jb}$. $K$ is an $n\times n$ matrix which in our case, it is chosen 
to be diagonal $K=M { 1\kern - .4em 1}$. Here, we used the notation in which
$e_{1,0}\equiv e_{N,1}$ and $e_{N,N+1}\equiv e_{1,N}$, that means
 the $(N+1)$-th layer is identified with the first one. In other words,
we are dealing with a compact discrete direction which could be considered
as a circle with circumference $R=Na$ with $a=M^{-1}$.\footnote{Using the notion of
distance in the non-commutative geometry, one can see that
the distance between the $(i+1)$-th and $i$-th layers is $a=M^{-1}$.
For recent discussion on the notion of distance in non-commutative
geometry see \cite{DJ}.}

A representation of any elements $f\in {\cal A}$ in $\cal H$ is
\be
\pi(f)=\sum_{i=1}^{N}f^i(x) e_{i,i}\, ,
\ee 
where $f^i(x):=f(x,y+ia)$ is a function on the manifold ${\cal M}_4$ defined 
at $i$-th layer and $y$ is the coordinate  of the discrete direction. Now,
we would like to study a gauge theory on this space. This
model corresponds to a four dimensional gauge theory with the gauge
group $SU(n)^N$ coupled to $N$ charged scalars which in general have quartic
potential.

Using the non-commutative formalizem of gauge theory introduced in the
previous section, we find the following expression for the gauge connection 
\bea
\pi(A)&=&\sum_{\alpha}\pi(g)_{\alpha}\;[D,\pi(f)_{\alpha}]
\cr &&\cr
&=&\sum_{i=1}^{N}\left[\gamma^{\mu}A^{i}_{\mu}\; e_{i,i}+\gamma^5 {M\over
\sqrt{2}}\;
(\phi^{i,i+1}e_{i,i+1}+\phi^{i,i-1}e_{i,i-1})\right]\, ,
\label{PIA}
\eea
where
\bea
A^i_{\mu}&=&\sum_{\alpha}g_{\alpha}^i\partial_{\mu}f_{\alpha}^i \, ,\cr
\phi^{i,i+1}&=&\sum_{\alpha}g_{\alpha}^i\;(f_{\alpha}^{i+1}-f_{\alpha}^i)\, ,\cr
\phi^{i,i-1}&=&\sum_{\alpha}g_{\alpha}^{i}\;(f_{\alpha}^{i-1}-
f_{\alpha}^{i})\, . 
\label{DIFG}
\eea
Similarly, one can also write down the representation of the curvature, 
$\pi(F)=\pi(dA)+\pi(A)^2$.
Plugging the result into the equation (\ref{ACT}), we can find the Yang-Mills 
action. Setting $U^{i,j}=\phi^{i,j}+1$, we get
\be
S_{YM}=\int d^4x\sum_{i=1}^N{\rm tr}\left[-{1\over 4g^2}F^i_{\mu\nu}
F^{i}_{\mu\nu}-{1\over 2}
f_s^2\overline{D_{\mu}U^{i,i+1}}D_{\mu}U^{i,i+1}+\cdots\right]\, ,
\label{YMAC4}
\ee
where
\bea
F^i_{\mu\nu}&=&\partial_{\mu}A_{\nu}^i-\partial_{\nu}A^{i}_{\mu}+[A^i_{\mu},
A^i_{\nu}]\, ,\cr&&\cr
D_{\mu}U^{i,i+1}&=&\partial_{\mu}U^{i,i+1}+A^i_{\mu}U^{i,i+1}-U^{i,i+1}
A^{i+1}_{\mu}\, ,\cr&&\cr
\overline{D_{\mu}U^{i,i+1}}&=&\partial_{\mu}U^{i+1,i}+A^{i+1}_{\mu}U^{i+1,i}-
U^{i+1,i}A^{i}_{\mu}\, ,
\eea
and $f_s^2=M^2/g^2$ with $g^2$ being the gauge coupling. The dots represent a
combination of the potential for the scalars $U^{ij}$ and the auxiliary fields
as well, which their forms
are not important for our
purpose.  Actually, as we already mentioned in the previous section, 
the auxiliary fields can be quotiented out. Alternatively, they can eliminated 
by their equations of motion, as they are not a dynamical field.
 \footnote{ For precise 
form of the auxiliary fields and their role in the
non-commutative geometry, the reader is referred to, for example, \cite{CHAM1}.}
Doing so, we will get a quartic potential for the scalars. These scalars
can get vacuum expectation values, and therefore this model would be 
equivalent to one considered in \cite{ACG} as a theory which dynamically 
generates fifth-dimension. Indeed, in the non-commutative geometry framework,
this fifth-dimension is nothing but the discrete dimension.

It is worth to note that as the distance between layers gets smaller and 
smaller, we will recover
a five -dimensional gauge theory with the gauge group $SU(n)$.
Physically, what we mean by $a\rightarrow 0$ is that, we are approaching
the IR limit where the energy scale of the theory is much smaller then the 
scale of the discrete dimension $g f_s$.
 To see this, 
we note that
in the non-commutative geometry the scalars $\phi^{i,j}$ play the role of the
gauge field in the discrete direction. To make this statement clear, we rewrite
the scalars as follow
\be\ba {lll}
M\phi^{i,i+1}&=\sum_{\alpha}g_{\alpha}^i\partial_{5}f^i_{\alpha}
&:=A^i_{5}\, ,\cr &&\cr
M\phi^{i,i-1}&=-\sum_{\alpha}g_{\alpha}^i{\bar \partial}_{5}f^i_{\alpha}
&:=-{\bar A}^i_{5}\, ,
\ea\ee 
where $\partial_5$ (${\bar \partial}_5$) is left (right) discrete derivative
\be
\partial_5\;f(y)=\frac{f(y+a)-f(y)}{a}, \;\;\;\;\;{\bar \partial}_5\;
f(y)=\frac{f(y)-f(y-a)}{a}\, .
\label{DISDER}
\ee
As $a\rightarrow 0$, these two derivatives become equal and therefore we get
$A_5^i={\bar A}^i_5$. Using this definition, the gauge connection (\ref{PIA}) 
reads
\be
\pi(A)=\sum_{i=1}^{N}\left[\gamma^{\mu}A^{i}_{\mu}\; e_{i,i}+{\gamma^5
\over \sqrt{2}} \left(
A^i_{5}\; e_{i,i+1}-{\bar A}^{i}_5\; e_{i,i-1}\right)\right]\, .
\label{PIA1}
\ee
Therefore, we get the following Yang-Mills action 
\be
S_{YM}=\int d^4x\sum_{i=1}^{N}{\rm tr}\left[-{1\over 4 g^2}F^i_{\mu\nu}
F^i_{\mu\nu}-{1\over 2g^2}{\bar F}^{i}_{\mu 5}F^i_{\mu 5}+ \cdots\right]\, ,
\label{ACT5}
\ee
here the dots represent the auxiliary field which can be integrated out.
In fact, if we had started with the corrected space of form, we would not have
seen the dots in the expression (\ref{ACT5}). Moreover
\bea
F^i_{\mu 5}&=&\partial_{\mu}A^i_5-\partial_5A^i_{\mu}+A^i_{\mu}A^{i}_5-
A^i_{5}A^{i+1}_{\mu}\, ,\cr &&\cr
{\bar F}^i_{\mu 5}&=&\partial_{\mu}{\bar A}^i_5-{\bar \partial}_5A^i_{\mu}+
A^i_{\mu}{\bar A}^{i}_5-
{\bar A}^i_{5}A^{i-1}_{\mu}\, .
\eea
Here we have applied the definition of the right and left discrete 
derivative to
$A^i_{\mu}$.

As $a\rightarrow 0$, we have $F^{i}_{\mu 5}={\bar F}^{i}_{\mu 5}$, and moreover
the summation can be replaced by an integral, more 
precisely we have $\sum_{i=1}^N \rightarrow {1\over a}\int_0^{Na}dy$, 
therefore the action (\ref{ACT5}) reads
\be
S_{YM}=\int d^4x\;dy \; {\rm tr}\left[-{1\over 4 g_{5}^2} F_{pq}F_{pq}
\right]\, ,
\ee
where 
\be
F_{pq}=\partial_{p}A_{q}-\partial_{q}A_{p}+[A_{p},
A_{q}],\;\;\;\;\;\;\;\;\; p,q=1,\cdots , 5
\ee
is the five-dimensional curvature and $g_{5}^{2}=Rg_{4}^2$ with $g_4=g/N$,
which is the gauge coupling of the diagonal subgroup of the original 
gauge group.

In order to find the Kaluza-Klein spectrum of the compactified five-dimensional
theory, we need the equation of motion of a massless scalar. In the
model we are considering, it is given by
\be
{\rm Tr}[\;D,[D,\pi(\psi)]\;]=0
\ee
where the trace is taken over both the Clifford algebra and the matrix 
structure. Setting $\psi_j=\varphi(x)\;exp(ik(y+ja))$, we find 
\be
g^{\mu\nu}\partial_{\mu}\partial_{\nu}\;\varphi(x)+({2\over a})^2\;
\sin^2({ka \over2})\;\varphi(x)=0\, .
\ee
Note that since the discrete direction is compact we have $k=2\pi l/Na$ for
$l=0, 1,\cdots , N$. Therefore, in the limit of $l\ll N$, we
recover precisely the correct Kaluza-Klein spectrum\footnote{The  
Kaluza-Klein spectrum of the four-dimensional theory with a discrete 
extra dimension has be also studied in \cite{HILL}.}     
\be
M_{KK}=\frac{2\pi l}{R}\, .
\ee
It can be also seen that in this limit, the five-dimensional Lorentz invariant 
is automatically restored. 


\section{Gravity Sector}

In this section we are going to introduce the gravity in our model. In fact one
advantage of looking at the model considered in \cite{ACG} from the 
non-commutative geometry point of view is that, in the framework of the
non-commutative geometry we will be able to formalize the theory of gravity
on the non-commutative space much similar to what we have for the gauge theory.
Although we shall only study the gravity sector, it is possible to 
have the gravity coupled to the gauge sector in the same time. We note that
the gravity in the non-commutative geometry has been studied in \cite{CHAM3}
as the gravity sector of Standard Model. Actually the content of this
 section is 
generalization of that in \cite{CHAM3} to 
the $N$-point space, though, our point of view is a little different.

Consider a space  which is taken to be a product of a
continuous four dimensional manifold times a discrete set of $N$ point. It is
very similar to what we had in the previous section, though, here we will
drop the assumption of the flatness of the four dimensional spacetime. Moreover
the distance between the layers is not taken to be constant.  
Nevertheless, the algebra
$\cal A$ has the same structure
\be 
{\cal A}={\cal C}^\infty({\cal M}_4) \otimes (\,\, \oplus_{i=1}^{N}
M_n({I \kern -.6em C})\,\,).
\label{ALG}
\ee

We would also like to introduce a local orthonormal basis for 
the cotangent bundle, $\Omega^1_{D}({\cal A})$. Here we use the following
convention for indices: the capital letters $A, B, \cdots$ run from ${\dot 1}$ 
to 
${\dot 5}$ \footnote{We use indices with dot for cotangent or tangent 
space in order not to be confused with spacetime indices.}
and the indices $a, b, \cdots$ run from ${\dot 1}$ to ${\dot 4}$. 
 The basis of the cotangent bundle, 
$\{e^{A}\}$, is
\be
\pi(e^a)=\sum_{i=1}^{N} \gamma^{a}\; e_{i,i}, \;\;\;\;\;\;
\pi(e^{\dot 5})=\sum_{i=1}^{N} {\gamma^{\dot 5}\over \sqrt{2}}
(\;e_{i,i+1}-e_{i,i-1}\;)\, ,
\label{BA}
\ee
with $\{\gamma^a,\gamma^b\}=2\eta^{ab}$ and $(\gamma^{{\dot 5}})^2=1$. 
The hermitian structure on $\Omega^1_{D}({\cal A})$ with the proper normalized
trace, Tr, is given by
\be
\langle \; e^{A}\; ,\; e^{B} \;\rangle={\rm Tr}\left(e^A\;(e^B)^*\right)=
\delta^{AB}\, ,
\label{HP}
\ee
which is essentially defined in terms of the Dixmier trace.

The Dirac operator can be chosen as follow
\be
D=\sum_{i=1}^{N}\left[\gamma^{a} e_{a}^{\mu}\partial_{\mu}\;e_{i,i}+
{\gamma^{\dot 5}\over \sqrt{2}}K_i\;(e_{i,i+1}+e_{i,i-1})\right]\,
\label{DICG}
\ee 
where $K_i=\lambda \phi_i(x) { 1\kern - .4em 1}$. This means that the distance
between the four-dimensional spaces is different. Nevertheless, we assume that
the expectation value of $\phi_i$ is constant of order one. 
Essentially, $\lambda$ 
plays the same role as $M$ in the previous section, in particular,
the length of the compact discrete direction is $R=N\lambda^{-1}$, which 
we shall assume to be fixed. Note that, $\{e^{\mu}_{a}\}$ in
(\ref{DICG}) is a vierbein, i.e.  an orthonormal basis of 
the section of the tangent bundle, so that
\be
e^{\mu}_{a }\;g_{\mu\nu }\;e^{\nu}_{b }=\eta_{ab},\;\;\;\;\;\;
e^{\mu}_{a}\;\eta^{ab}\;e^{\nu}_{b }=g^{\mu\nu}. 
\ee

Suppose $\rho=\sum_{\alpha}g_{\alpha}\;df_{\alpha}$ be a one form, i.e.
$\rho\in \Omega^1_{D}({\cal A})$, using the Dirac operator (\ref{DICG}),
 we get
\be
\pi(\rho)=\sum_{i=1}^{N}\left[\gamma^{a}e_{a}^{\mu}\rho_{\mu i}\; e_{i,i}+
 {\gamma^{\dot 5}\over \sqrt{2}}\lambda 
\phi_i(x)\;\left(\rho_{5 i}\; e_{i,i+1}-{\bar \rho}_{5 i}\; e_{i,i-1}\right)
\right]\, ,
\label{ONE}
\ee
where $\rho_{\mu i}, \rho_{5 i}, {\bar \rho}_{5 i}$ are defined the same as
those in (\ref{DIFG}), of course with a different sign for 
${\bar \rho}_{5 i}$. Using the spacetime gamma matrices, $\gamma^{\mu}=
\gamma^ae^{\mu}_a,\;\gamma^{5}_i=\gamma^{{\dot 5}}e^5_{{\dot 5}i}$ with
$e^5_{{\dot 5}_i}=\phi_i(x)$, the expression of the one-form (\ref{ONE}) can
be recast as  
\be
\pi(\rho)=\sum_{i=1}^{N}\left[\gamma^{\mu}\rho_{\mu i}\; e_{i,i}+
{\gamma^{5}_i \over \sqrt{2}}\lambda
\;\left(\rho_{5 i}\; e_{i,i+1}-{\bar \rho}_{5 i}\; e_{i,i-1}\right)\right]\, ,
\label{ONE1}
\ee
which is essentially analogous to (\ref{PIA1}) for non-flat space. It can also
be shown that 
the expression of $\pi(d\rho)$ modulo the auxiliary fields is
\bea
\pi(d\rho)&=&\sum_{i=1}^{N}\left[\gamma^{\mu\nu}\partial_{\mu}
\rho_{\nu i}\; e_{i,i}+{\gamma^{\mu 5}_i\over \sqrt{2}}\lambda
 \left(\partial_{\mu}
\rho_{5 i}+\rho_{\mu i}-\rho_{\mu i+1}\right)\;e_{i,i+1}  \right. \cr
 &&\left. \right. \cr
&-&\left. {\gamma^{\mu 5}_i\over \sqrt{2}}\lambda \left(\partial_{\mu}
{\bar \rho}_{5 i}+\rho_{\mu i-1}-\rho_{\mu i}\right) \; e_{i,i-1}\right]\, .
\label{TWO}
\eea

A connection, $\bigtriangledown$, on $\Omega^1_{D}(\cal A)$ is defined by
$\bigtriangledown e^A=-\omega^{AB} \otimes e^B$ with $\omega^{AB}\in
\Omega^1_{D}(\cal A)$.  Using equation (\ref{ONE1}), it can be seen that
$\pi(\bigtriangledown)$ in the basis $\{e^A\}$ has following general form 
\be
\pi(\omega^{AB})=\sum_{i=1}^{N}\left[\gamma^{\mu}\omega^{AB}_{\mu i}\; 
e_{i,i}+
{\gamma^5_i\over \sqrt{2}}\lambda \left(\chi^{AB}_{i}\;e_{i,i+1}-
{\bar \chi}^{AB}_{i}\;e_{i,i-1} \right)\right] \, ,
\ee
>From the hermiticity property of $\bigtriangledown$ we have 
\be
\omega^{AB}_{\mu i}=-\omega^{BA}_{\mu i},\;\;\;\;\;\;\;
\chi^{AB}_{i}=-{\bar \chi}^{BA}_{i}\, .
\ee
The components of the torsion and Riemann curvature defined by
$T^A=T(\bigtriangledown)\;e^A$ and ${\cal R}(\bigtriangledown)e^A=
{\cal R}^{AB}\otimes ^B$ respectively, are given by \cite{CHAM3}
\bea
T^A&=&\pi(de^A)+\pi(\omega^{AB})\, \pi(e^B) \, ,\cr
&&\cr
{\cal R}^{AB}&=&\pi(d\omega^{AB})+\pi(\omega^{AC})\pi(\omega^{CB})\, .
\eea

Using the most general expression of the one- and two-form, (\ref{ONE1}) and
(\ref{TWO}), the components of the torsion and curvature can be written as
follow
\bea
T^a&=&\sum_{i=1}^{N}\left[\gamma^{\mu\nu}\left(\partial_{\mu}e^a_{\nu}+
\omega^{ab}_{\mu i}e^b_{\nu}\right)\;e_{i,i}
+{\gamma^{\mu 5}_i\over \sqrt{2}}\left(\omega^{a {\dot 5}}_{\mu i}
e^{\dot 5}_{5 i}-\lambda \chi^{ab}_{i}e^b_{\mu}\right)\,e_{i,i+1}
\right.\cr &&\left. \right.\cr
&-&\left. {\gamma^{\mu 5}_i\over \sqrt{2}}\left(\omega^{a {\dot 5}}_{\mu i}
e^{\dot 5}_{5 i}-\lambda {\bar \chi}^{ab}_{i}e^b_{\mu}\right)\,e_{i,i-1}\right]
\, ,\cr &&\cr
T^{\dot 5}&=&\sum_{i=1}^{N}\left[\gamma^{\mu\nu}
\omega^{{\dot 5}b}_{\mu i}e^b_{\nu}\;e_{i,i}
+{\gamma^{\mu 5}_i\over \sqrt{2}}\left(\partial_{\mu}
e^{\dot 5}_{5 i} -\lambda \chi^{{\dot 5}b}_{i}e^b_{\mu}\right)\,e_{i,i+1}
\right.\cr &&\left. \right.\cr
&-&\left. {\gamma^{\mu 5}_i\over \sqrt{2}}\left(\partial_{\mu}e^{\dot 5}_{5 i}
-\lambda {\bar \chi}^{ab}_{i}e^b_{\mu}\right)\,e_{i,i-1}\right]
\eea
for the torsion, and for the curvature we find
\bea
{\cal R}^{AB}&=&\sum_{i=1}^{N}\left[{1\over 2}
\gamma^{\mu\nu} {\cal R}^{AB}_{\mu\nu i}\,
e_{i,i}+{\gamma^{\mu 5}_i \over \sqrt{2}}\lambda 
 \left( {\cal Q}^{AB}_{\mu i} e_{i,i+1}
-{\bar {\cal Q}}^{AB}_{\mu i} e_{i,i-1}\right)\right]\, ,
\label{RCUR}
\eea
where
\bea
{\cal R}^{AB}_{\mu\nu i}&=&\partial_{\mu}\omega^{AB}_{\nu i}
-\partial_{\nu}\omega^{AB}_{\nu i}
+\omega^{AC}_{\mu i}\;\omega^{CB}_{\nu i}-
\omega^{AC}_{\nu i}\omega^{CB}_{\mu i}\, ,\cr
&&\cr
{\cal Q}^{AB}_{\mu i}&=&\partial_{\mu}\chi^{AB}_{i}
+\omega^{AB}_{\mu i}-\omega^{AB}_{\mu i+1}
+\omega^{AC}_{\mu i}\;\chi^{CB}_{i}-
\chi^{AC}_{i}\omega^{CB}_{\mu i+1}\, ,\cr
&&\cr
{\bar {\cal Q}}^{AB}_{\mu i}&=&\partial_{\mu}{\bar \chi}^{AB}_{i}
+\omega^{AB}_{\mu i}-\omega^{AB}_{\mu i-1}
+\omega^{AC}_{\mu i}\;{\bar \chi}^{CB}_{i}-
{\bar \chi}^{AC}_{i}\omega^{CB}_{\mu i-1}\, .
\eea

Finally the Einstein-Hilbert action is
\bea
S_{EH}&=&\kappa^{-2}\langle {\cal R}^{AB}e^B\; ,\; e^A 
\rangle\cr &&\cr
&=&\kappa^{-2}\int d^4 {\rm Tr}\left({\cal R}^{AB} e^B 
(e^A)^*\right)\, .
\label{EHA}
\eea
>From the equation (\ref{BA}), (\ref{HP}) and (\ref{RCUR}), the 
Einstein-Hilbert action, (\ref{EHA}), reads
\be
S_{EH}=\int \sqrt{\det(g)} d^4x\;\sum_{i=1}^{N}
\left[{1\over \kappa^2}
 e^{\mu}_{a}e^{\nu}_{b} {\cal R}^{ab}_{\mu\nu i}+{\lambda\over 2 \kappa^2}
 e^{5}_{{\dot 5} i} e^{\mu}_{a}\left(
 {\cal Q}^{a {\dot 5}}_{\mu i}+{\bar {\cal Q}}^{a {\dot 5}}_{\mu i}
 -{\cal Q}^{{\dot 5} a}_{\mu i}-{\bar {\cal Q}}^{{\dot 5} a}_{\mu i}\right)
 \right]\, .
\label{EHAE}
\ee

One can now impose the torsionless condition which leads to the following 
conditions:  
 $\omega^{ab}_{\mu i}=\omega^{ab}_{\mu}$ for all $i$, and $\chi^{AB}_i
={\bar \chi}_i^{AB}$. Moreover, we get
\be
\partial_{\mu} e^{{\dot 5}}_{5 i}=\lambda \chi^{{\dot 5} b}_{i}e^{b}_{\mu}\, .
\ee
Plugging these conditions into the (\ref{EHAE}), 
one finds the following action for the gravity 
\be
S_{EH}=\int \sqrt{\det(g)} d^4x \left[{1\over \kappa^2_4}e^{\mu}_a e^{\nu}_{b}
 {\cal R}^{a b}_{\mu \nu}-{1\over 2}\sum_{i=1}^{N}
\partial_{\mu}\sigma_i\partial^{\mu}\sigma_i\right]\, ,
\ee
where $\phi_i(x)=e^{-\kappa \sigma_i(x)/2}$ and $\kappa^2_4=\kappa^2/N$. 
As a conclusion, the Einstein-Hilbert
action for the non-commutative space given by (\ref{ALG}), turns out to be
the gravity action plus $N$ scalars. In order to understand the role of these
scalars one has to consider the gravity coupled to the Yang-Mills sector.
For this purpose, one can write the Yang-Mills action in the same way as we 
did in the previous section, but with the Dirac operator (\ref{DICG}).
Of course, this in not what we are going to do now, we would rather to consider the
case where $\lambda \rightarrow \infty$. Physically, this corresponds to the 
limit 
where the good description would be in the terms of a five-dimensional gravity,
much similar to what we had in the Yang-Mills sector when $M\rightarrow 
\infty$.

Using the notation of the left and right discrete derivative (\ref{DISDER}),
 and
setting $\lambda \chi_i^{AB}=\omega^{AB}_{5 i}$ and
$\lambda {\bar \chi}_i^{AB}={\bar \omega}^{AB}_{5 i}$, we find
\bea
\pi(\omega^{AB})&=&\sum_{i=1}^{N}\left[\gamma^{\mu}\omega^{AB}_{\mu i}\; 
e_{i,i}+
{\gamma^5_i\over \sqrt{2}} \left(\omega^{AB}_{5i}\;e_{i,i+1}-
{\bar \omega}^{AB}_{5i}\;e_{i,i-1} \right)\right] \cr &&\cr
{\cal R}^{AB}&=&\sum_{i=1}^{N}\left[{1\over 2}
\gamma^{\mu\nu} {\cal R}^{AB}_{\mu\nu i}\,
e_{i,i}+{\gamma^{\mu 5}_i \over \sqrt{2}} 
 \left( ({\cal R}^{AB}_{\mu 5 i}-{\cal L}) e_{i,i+1}
-({\bar {\cal R}}^{AB}_{\mu 5i}-{\bar {\cal L}}) e_{i,i-1}\right)\right]\, ,
\eea
where ${\cal L}=\lambda^{-1}\omega^{AC}_{5 i}\partial_5\omega^{CB}_{\mu i},\;\;
{\bar {\cal L}}=\lambda^{-1}{\bar \omega}^{AC}_{5 i}{\bar \partial}_5
\omega^{CB}_{\mu i}$ and 
\bea
{\cal R}^{AB}_{\mu 5 i}&=&\partial_{\mu}\omega^{AB}_{5 i}
-\partial_{5}\omega^{AB}_{\mu i}
+\omega^{AC}_{\mu i}\;\omega^{CB}_{5 i}-
\omega^{AC}_{5 i}\omega^{CB}_{\mu i}\, ,\cr &&\cr
{\bar {\cal R}}^{AB}_{\mu 5 i}&=&\partial_{\mu}{\bar \omega}^{AB}_{5 i}
-{\bar \partial}_{5}\omega^{AB}_{\mu i}
+\omega^{AC}_{\mu i}\;{\bar \omega}^{CB}_{5 i}-
{\bar \omega}^{AC}_{5 i}\omega^{CB}_{\mu i}\, .
\eea
Keeping in mind that in the limit of $\lambda \rightarrow \infty$ 
we have
$\omega^{AB}_{\mu 5 i}={\bar \omega}^{AB}_{\mu 5 i}$, the 
action (\ref{EHA}) reads
\be
S_{EH}=\int \sqrt{\det(g)} d^4x\;\sum_{i=1}^{N}
\left[{1\over \kappa^2}
 e^{\mu}_{a}e^{\nu}_{b} {\cal R}^{ab}_{\mu\nu i}+{1\over \kappa^2}
 e^{5}_{{\dot 5} i} e^{\mu}_{a}\left(
 {\cal R}^{a {\dot 5}}_{\mu 5 i}+{\bar {\cal R}}^{a {\dot 5}}_{\mu 5i}
 \right)
 \right]\, .
\ee
Here we have dropped those terms which  are proportional 
to $\lambda^{-1}$.  In the
limit where $\lambda \rightarrow \infty$, 
one has also to replace the summation with an integral. 
Doing so, we find the five-dimensional gravity
action as following
\be
S_{EH}=\kappa_5^{-2}\int \sqrt{\det(G)}\, d^4x\, dy\, e^p_A e^q_B 
{\cal R}^{AB}_{pq}\, ,
\ee
where ${\cal R}^{AB}_{pq}, \det(G)=\phi^2(x,y)\det(g)$ 
and $\kappa_5^2=R\kappa_4^2$ are 
five-dimensional curvature, metric and Newton constant, respectively. 

Note that, in the geometry we are considering for our spacetime, by dimensional 
reduction from five dimensions to four dimensions we will not get a gravity
 coupled to gauge field, as we used to get in the ordinary Kaluza-Klein 
reduction, where the $G_{\mu 5}$ plays the role of the gauge field in the 
four-dimensional theory. In fact, in our case the four-dimensional Yang-Mills 
and
gravity sectors are coming from the five-dimensional Yang-Mills and
gravity sectors, respectively. As we shall discuss in the next section, the
effects of the five-dimensional gravity in the four-dimensional Yang-Mills 
sector will be appeared in the potential for the scalars in the Yang-Mills 
sector.

\section{Conclusions}

In this note we considered the Yang-Mills theory as well as the gravity in a
non-commutative space given by a four-dimensional manifold times a set of 
discrete
 $N$-points. The Yang-Mills theory which we studied in this paper is 
a gauge theory with gauge group $SU(n)^N$ coupled to $N$-charged scalars.
In the high energy limit the theory is a four dimensional field theory,
while in the IR limit, the theory behaves like a five-dimensional gauge
theory with the gauge group $SU(n)$.
The same as that in \cite{ACG}, one can think about this procedure as a
dynamically generating dimension. 

This model can also be thought as a latticization of a five-dimensional 
gauge theory. From the non-commutative geometry point of view, this can be
seen by noting that the non-commutative five-dimensional spacetime we
have used so far, can be considered as a 
five-dimensional space, $(x_i,y),\;i=1,\cdots 4$, with following 
non-commutative relation
\be
[y,dy]=ady\, ,
\ee
where $a$ is a constant. All other coordinates commute. Moreover, we have
\be
df(y)=dy\, (\partial_y f)(y)=({\bar \partial}_y)(y)\,dy \, , 
\ee
where the left and right derivatives are defined as (\ref{DISDER}).
One can show that a gauge theory on this space
will be a five-dimensional gauge theory latticized in one dimension, much
similar to (\ref{YMAC4}) (see for example \cite{MULLER}).

An advantage working in the framework of the non-commutative
is that the gravity can be added to the game in the same way as the
Yang-Mills sector.
Although, in this paper we have only considered the gravity and Yang-Mills 
sectors separately, one could consider both of them at the same time. 
In particular, in this case we will get the following potential for
the scalars in the Yang-Mills sector (see also \cite{LMM})
\be
V_i\sim (|U_{i,i+1}|^2-e^{-\kappa \sigma})^2.
\label{pot}
\ee
Here we set $\phi_i(x)=\phi(x)=e^{-\kappa \sigma/2}$. Furthermore, 
the Kaluza-Klein spectrum for this case is
\be
M^2_{{\rm KK}}=e^{-{\kappa \sigma }}\;({2\over a})^2\sin^2({ka\over 2}).
\ee
Note that, as we can see from 
(\ref{pot}), the effect of the five-dimensional gravity in four-dimensional
theory is given by a potential in the from of that in Randall-Sundrum 
model \cite{RS}. In fact, one could think about the theory
we are considering here, as $N$-copies of the Randall-Sundrum model.
One could also add the fermions to the theory. 

We would like to note that, as far as the five-dimensional theory
concerns, there is no difference between a theory with parameters $(a, N)$ and
$(a', N')$, provided of course $aN=a'N'$. Nevertheless, as we are approaching 
the UV limit we will end up with two different four-dimensional theories. For
example, If we started with a five-dimensional gauge theory with gauge group
$SU(n)$, the four-dimensional theories would be either $SU(n)^N$ 
gauge theory with parameters $g$ and $f_s$ or $SU(n)^{N'}$ gauge theory 
with parameters $g'$ and $f'_s$. Moreover, we have the following relations
between the parameters of these two description
\be
g'={N'\over N}g,\;\;\;\;\;\;\;\;f_s=f'_s\; .
\ee
Therefore, it seems that starting from a five-dimensional gauge theory we
will have several UV completions. One might suspect that in the framework of the
non-commutative geometry, these issue would be related to the notion of the
``Morita equivalence''. 
It would be nice to see if we can make this relation more precise. We hope to 
come back to this point in the future.  

We also hope that the interpretation we have made here, 
could be used for the further 
study of the (de)constructing dimensions story.

{\bf Acknowledgments}: I would like to thank N. Arkani-Hamed for comments
and correspondence.

\end{document}